\documentclass[
    journal,
    twocolumn,
    ]{IEEEtranTCOM}
\usepackage{graphicx}
\usepackage{amsmath}
\usepackage{amssymb}
\usepackage{amsthm}
\usepackage{afterpage}
\usepackage{verbatim}
\usepackage{psfrag}
\usepackage{color}
\usepackage{cite}
\usepackage{setspace}
\usepackage{epsfig}
\usepackage{epstopdf}
\usepackage{footmisc}
\usepackage{fmtcount}
\usepackage{stackrel}
\usepackage{float}
\usepackage{breqn}
\usepackage{enumerate}
\usepackage{graphicx}
\usepackage{subcaption}
\usepackage{graphicx}
\DeclareMathOperator*{\argmax}{arg\,max}
\DeclareMathOperator*{\argmin}{arg\,min}
\begin{document}
%\doublespace
\title{Outlier Detection and Optimal Anchor Placement for 3D Underwater Optical Wireless Sensor Networks Localization}
\author{Nasir Saeed,~\IEEEmembership{Member,~IEEE},  Tareq Y. Al-Naffouri,~\IEEEmembership{Member,~IEEE}, Mohamed-Slim Alouini,~\IEEEmembership{Fellow,~IEEE}
\thanks{This work is
supported by Office of Sponsored Research (OSR) at King Abdullah University of Science and Technology (KAUST). 

The authors are with the Computer Electrical and Mathematical Sciences \& Engineering (CEMSE) Division, KAUST, Thuwal, Makkah Province, Kingdom of Saudi Arabia, 23955-6900. This work is an extension to our previous work in \cite{Nasir2018spawc}.}
}
\maketitle
\begin{abstract}
Location is one of the basic information required for underwater optical wireless sensor networks (UOWSNs) for different purposes such as relating the sensing measurements with precise sensor positions, enabling efficient geographic routing techniques, and sustaining link connectivity between the nodes. Even though various two-dimensional UOWSNs localization methods have been proposed in the past, the directive nature of optical wireless communications and three-dimensional (3D) deployment of sensors require to develop 3D underwater localization methods. Additionally, the localization accuracy of the network strongly depends on the placement of the anchors.
Therefore, we propose a robust 3D localization method for partially connected UOWSNs which can accommodate the outliers and optimize the placement of the anchors to improve the localization accuracy. The proposed method formulates the problem of missing pairwise distances and outliers as an optimization problem which is solved through half quadratic minimization. Furthermore, analysis is provided to optimally place the anchors in the network which improves the localization accuracy. The problem of optimal anchor placement is formulated as a combination of Fisher information matrices for the sensor nodes where the condition of D-optimality is satisfied. The numerical results indicate that the proposed method outperforms the literature substantially in the presence of outliers. 
\end{abstract}
\IEEEpeerreviewmaketitle
\section{Introduction}
The deployment of underwater wireless sensor networks (UWSNs) opens up a wide range of possible applications as the large percentage of Earth's surface is covered by water in the form of oceans, seas, and rivers. UWSN applications include but not limited to assisted navigation, disaster management, offshore exploration, ocean sampling, and underwater monitoring \cite{AKYILDIZ2005, Akyildiz2014}. However, connecting underwater sensors is a challenging task due to the hostile underwater wireless channel conditions \cite{celik2018modeling}.  Radio frequency (RF), acoustic, and optical systems are three main forms of underwater wireless communications and each has virtues and drawbacks. Underwater RF systems can provide average speed data rate, however, they can only operate at surface levels and cannot operate at higher depths due to significantly high absorption. Therefore,  acoustic channels are mostly preferred for UWSNs due to its low absorption and long transmission distance. Nonetheless, acoustic systems suffer from bandwidth scarcity and low propagation speed (i.e., 1500 m/s) which yields a considerable latency at large distances. To overcome the problem of low data rate and high latency acoustic signals, underwater optical wireless communication (UOWC) has recently attracted attention by its high data rates and low latency levels \cite{Zeng2017}. Nevertheless, the low transmission range of UOWC systems necessitates practical networking and control solutions to realize underwater optical wireless sensor networks (UOWSNs) in real life.

The unique properties of UOWC also necessitate for an innovative reexamination of the localization problem as it is also an important task for UOWSNs to geographically tag the data,  mitigate the limited transmission range via multi-hop routing based on sensor locations, and to sustain the link connectivity and quality via pointing, acquisition, and tracking (PAT) mechanisms. Arguably, UOWSN localization is a far more challenging task than terrestrial localization due to the non-availability of global positioning system (GPS) signals and distinct propagation characteristics of light beams in aquatic medium \cite{Akhoundi2017underwater}. In addition, two-dimensional localization methods can work well for terrestrial networks but the directivity of optical beams and 3D deployment of sensors in  the underwater environment require to develop  three-dimensional (3D) localization methods which are more challenging due to the sparsity of UOWSN deployment  and hostile underwater environment. Besides that, the number of anchor nodes (beacons) is limited for UOWC in comparison to terrestrial communication. All of these considerations make the 3D localization of UOWSNs a significant task where very few choices are available.

Besides the aforementioned challenges,  the limited transmission range and other limiting factors of underwater optical channel such as absorption, scattering, turbulence, salinity, and air bubbles leads to develop a multi-hop network with outliers. Therefore for a centralized localization scheme, it is not only required to estimate all the pairwise distances but also to remove the outliers. Although a number of conventional matrix completion methods have been proposed in the past to estimate the missing pairwise distances, for example, singular value thresholding (SVT) \cite{Cai2010}, atomic decomposition for minimum rank approximation (Admira) \cite{ Lee2009}, Optspace \cite{Keshavan2010}, augmented Lagrange multiplier method \cite{Lin2013, Wang2016}, and alternating minimization \cite{Gamarnik2016}, all of these matrix completion methods only estimate the missing elements of a partially filled matrix and do not consider the outliers, and are therefore not robust to the outliers. 

Additionally, placement of anchors for localization is a non-trivial task and is of great interest to the localization research community. The literature on the problem of optimal anchor placement mainly considers a two-dimensional scenario where a single node is to be localized with a minimum localization error \cite{Zhang1995, Bishop20071,Bishop20072,Bishop2009, Hamdollahzadeh2016,Saeed2017L, Rusu2017}. Three-dimensional optimal placement of anchors for a single vehicle localization is investigated in \cite{Ramirez2013}. The analytical characterization of optimal sensor placement for three-dimensional scenarios with multiple sensor nodes is still an open research problem. 

In this paper, we propose a robust 3D localization method for UOWSNs with limited connectivity. The problem of network localization seeks to find the position of each node given that few anchor nodes and some of the noisy inter-node distances are available. Because of the limited UOWC transmission distance, we consider a multi-hop UOWC system to extend the communication range. Furthermore, it is crucial to estimate the missing distances accurately since the inter-node noisy distances are not always available for all node pairs. Also, the severe UOWC channel conditions can cause outliers to some of the inter-node distances. Hence, it is important to remove the outliers otherwise the outliers can propagate the error throughout the network and yield a large localization error. Additionally, we provide an analytical expression for the optimal placement of anchors which reduces the network localization error.

The main contributions of the paper can be summarized as follows: 
\begin{itemize}
\item 
Two-dimensional localization methods for UOWSNs have been studied in \cite{Akhoundi2017underwater, Saeed2017, Nasir2018limited}. To the best of our knowledge, this paper is first to consider a 3D localization for UOWSNs.

\item 
The directive nature and limited transmission distance of UOWC lead to a partially connected network where there are many missing inter-node distances which are required to be accurately estimated. Hence, we develop a low-rank matrix approximation method which can accurately estimate the missing inter-node distances.
\item 
Some of the inter-node distance can have a large error and introduces outliers to which the conventional 3D network localization methods are quite susceptible. Consequently, a closed-form convergent iterative solution is proposed which can accommodate the outliers.

\item An analytical optimal condition is provided for the anchor nodes to satisfy which maximizes the localization accuracy for multiple sensor nodes.
\end{itemize}
The remainder of the paper is organized as follows. In Section \ref{sec:related} the literature on UOWSNs localization, matrix completion, and optimal anchor placement is presented. In Section \ref{systmodel}, we introduce the network model and define the single hop pairwise distance measurements for UOWSNs. Section \ref{sec:proposed} presents the closed form solution for the proposed  3D localization method where the missing pairwise distances are estimated and outliers are removed. Furthermore, Section \ref{sec:proposed} also presents the optimal placement of anchors in  the 3D underwater environment to improve the localization accuracy.
Section \ref{results} provide the numerical results to validate the performance of the proposed method. Finally, Section \ref{conc} concludes the paper with a few remarks.

\section{Related Work}\label{sec:related}
Localization of nodes in UOWSNs is of great importance which can enable numerous applications. Conventionally, localization methods are either range based or range free where the range based methods rely on different ranging methods to estimate the distances, and then, estimate position of the node based on the estimated distances. The range-free localization methods provide coarse position estimation where usually the area containing the node is estimated. Since range-free localization methods are not yet developed for UOWSNs, we focus on range based methods.

The  two-dimensional range-based localization methods for UOWNs can be classified into two categories as distributed and centralized methods. The authors in \cite{Akhoundi2017underwater} have proposed for the first time a recieved signal strength (RSS) and time of arrival (ToA) based distributed localization method. The authors have considered an optical base station (OBS) placed in a hexagonal cell which serves as an anchor node for the users.  The users are able to estimate the distance to multiple OBSs and then estimate its position by using linear least square estimation. A centralized hybrid acoustic and optical RSS ranging based localization method have been proposed in \cite{Saeed2017} where the authors have considered a weighting strategy to give more importance to accurate ranging measurements. A centralized RSS based localization method has also been proposed in \cite{Nasir2018limited}, where the nodes estimate and forward the single hop RSS based distances to the centralized node. The centralized node is then able to estimate the position of each node. However, all of the above UOWSNs localization methods are two-dimensional and do not consider the 3D nature of the underwater environment. We refer the interested reader to \cite{Saeed2018Survey} which presents a comprehensive survey on UOWSNs localization. 

As the centralized localization methods are based on dimensionality reduction techniques which require all  of the pairwise distance estimations between the nodes, robust matrix completion methods are needed to estimate the missing pairwise distances and remove the outliers. A number of researchers have provided theoretical constraints about the matrix rank, sampling scheme, and missing rate to effectively complete a matrix in low rank. In general, matrix completion schemes can be classified into three classes. The first class consists of matrix factorization based schemes where the missing entries of a partially filled matrix of size $M \times N$ and rank $r$ are estimated from the product of two matrices $M \times r$ and $N \times r$ \cite{Nathan2004, Shen2014}. Matrix completion schemes based on matrix factorization are non-convex and they require prior information about the rank of the matrix. The authors in \cite{Wen2012} have dynamically adjusted the rank of the matrix to estimate the missing elements via matrix factorization. The matrix completion problem in\cite{Wen2012} is formulated as a non-linear successive over relaxation  problem which is solved by using linear least square method.

The second class of matrix completion schemes consists of nuclear norm minimization methods which are convex. The authors in \cite{Cai2010} proposed a singular value thresholding  (SVT) method to estimate the missing entries of a partially filled matrix. In \cite{Lin2013, Wang2016},  the authors have applied an augmented Lagrange multiplier (ALM) method for nuclear norm minimization which pads the missing entries of the matrix with zeroes and considers the completed matrix as a sum of true complete matrix and error matrix. The alternating direction (AD) method is proposed in \cite{Gamarnik2016} to minimize the nuclear norm which is similar to the method proposed in \cite{Lin2013, Wang2016} but the error matrix is not explicit. In \cite{Hu2013}, the authors have proposed truncated nuclear norm (TNN) minimization which is a special case of SVT where the largest singular values are equal to zero. The third class of matrix completion schemes is based on manifold optimization methods. The authors in \cite{Keshavan2010} have proposed a low-rank matrix completion method (also called Optspace) by minimization over the Grassman manifolds \cite{Keshavan2010}. A low-rank matrix completion method based on manifold optimization is also proposed in \cite {Bart2013} which uses the Riemannian manifold. 
However, all of these matrix completion methods only estimate the missing elements of a partially filled matrix and do not consider the outliers, and  therefore, are not robust to the outliers.

Outliers can significantly contaminate the ranging measurements, and thus lead to large localization error. A number of outlier detection techniques have been presented in \cite{zhang2010outlier} for data mining, predictive modeling, cluster analysis, and wireless sensor networks. Recently, outlier rejection methods have been developed for multilateration based iterative localization \cite{Wang2008}. However, in sparse networks, each node may not be directly connected to the anchors and thus cannot perform multilateration for localization. Hence, networks such as  UOWSNs with sparsity and hostile underwater environment (absorption, scattering, geometrical losses, turbulence, and air bubbles) require to develop robust localization scheme to encounter the outliers produced by all these phenomena. Therefore, in this paper, we have developed a novel 3D localization algorithm which is not only robust to the ranging errors but also accounts for such outliers.

In addition to the presence of outliers, UOWSNs require to develop 3D localization methods which are more applicable for UOWSNs. Until now, 3D localization methods for UOWSNs do not exist, however, there exists a number of 3D localization methods for underwater acoustic sensor networks (UASNs). The authors in \cite{Teymorian2009} have proposed for the first time a 3D localization scheme for UASNs where a projection based method is used to project the 3D localization problem into its 2D counterpart. The authors in \cite{guo2013localization} have proposed an anchor-free 3D localization scheme for UASNs where the anchors are able to float along the network. The sensor node computes the ranges to a mobile anchor and computes its 3D location. A 3D multihop localization method has been proposed in \cite{ren2014orthogonal} where multilateration is used by the sensor nodes to compute their position by getting the measurements from at least three anchor nodes. A top-down approach has been used in \cite{zhang2014top} for UASNs localization where the sensor nodes near to the surface buoys (anchors) estimate their positions and then act as anchors. Similarly, a 3D localization method has been proposed in \cite{uddin2016low} with a small number of anchors where once a sensor node estimates its location it becomes a new anchor.  Recently, a 3D localization method for UASNs have been proposed in \cite{mridula2018localization} where the uncertainty in the anchor positions is considered.

Furthermore, the placement of anchors for network localization methods is also an important and challenging problem. The literature on optimal placement of anchors consists of two types of mathematical formulations. One is parameter optimization and the other is optimal control. The optimal control based formulation is usually adopted for path planning problems \cite{Xu2003, Jarurat2007} while parameter optimization is mainly used for optimal placement of anchors \cite{Zhang1995, Bishop20071,Bishop20072,Bishop2009, Hamdollahzadeh2016,Saeed2017L, Rusu2017}. Parameter optimization method estimates the position of sensors such that the uncertainty in position estimation is minimized. The uncertainty in position is characterized by Fisher information matrix (FIM) which is the inverse of Cramer Rao lower bound (CRLB). Any unbiased estimator which can achieve the CRLB is efficient, and therefore, the position of anchors which maximizes the determinant of FIM for the position estimation of all nodes is said to be the optimal position of anchors.

The proposed localization method for UOWSNs is fundamentally different and practical than the aforementioned schemes where we have tackled multiple problems such as estimating the missing distances, removal of outliers, and optimizing the anchor positions to improve the localization accuracy.  The derived expressions for the proposed 3D localization are general enough to be applicable for any multi-hop wireless communication network (including terrestrial and underwater wireless communication systems). However, we especially focused on multi-hop UOWSNs since the UOWC channel is more susceptible to the outliers caused by the intrinsic properties of optical light and underwater aquatic medium. Before going into the details of the proposed localization scheme, we formally introduce the network model.
\begin{figure}
\centering
\includegraphics[width=0.7\columnwidth]{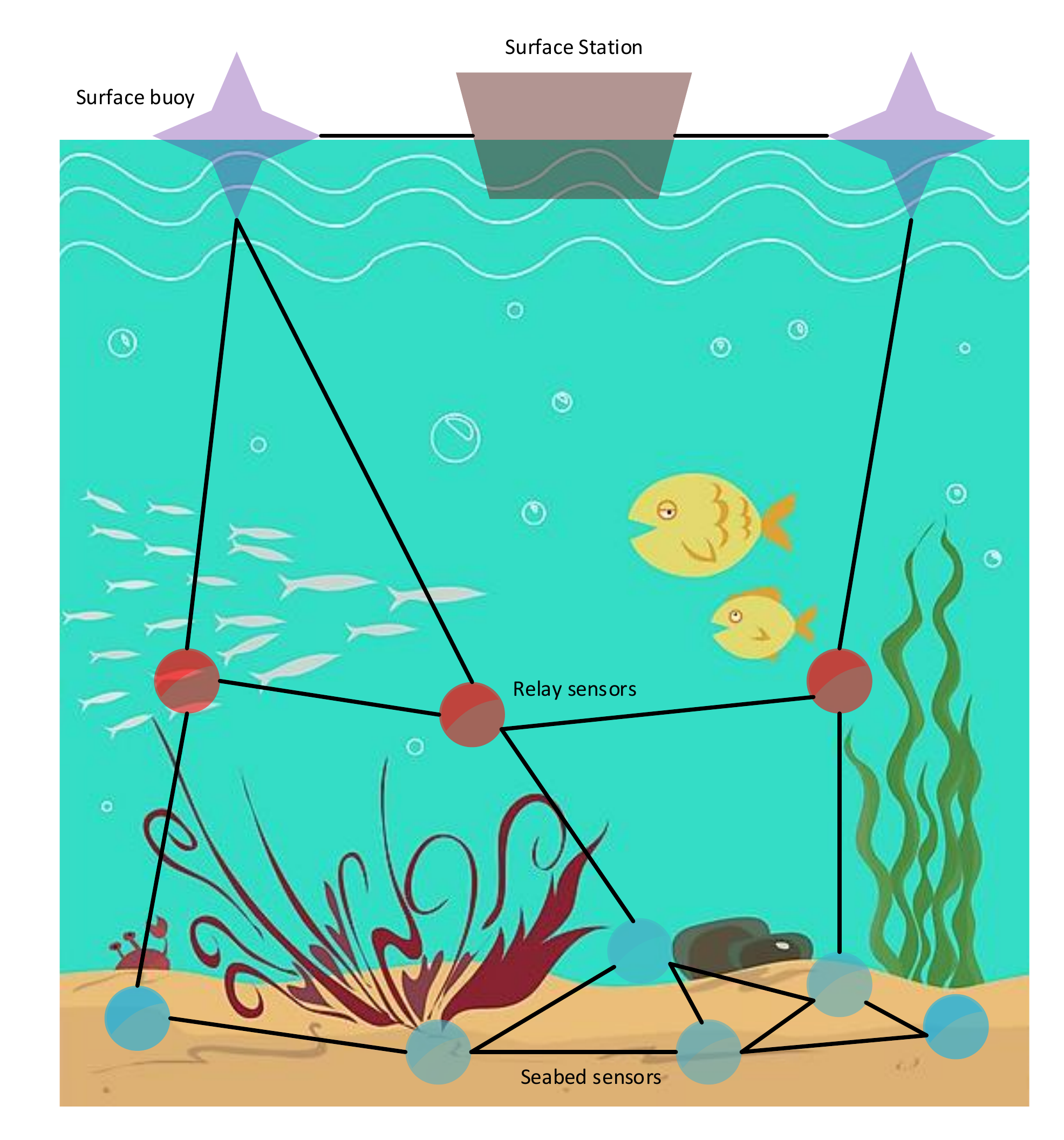}  
\caption{3D UOWSN model.\label{fig:systemmodel}} 
\vspace{-1.5 em}
\end{figure}
\section{Network Model}
\label{systmodel}
Consider a 3D UOWSN architecture as shown in Fig.~\ref{fig:systemmodel} where the sensor nodes are deployed at the seabed and their information is delivered to the surface buoys via relay sensors in a multi-hop fashion.  It is assumed that the location of surface buoys are known and they act as anchor nodes in the network.  To introduce the notations for $m$ seabed sensors, $n$ relay sensors, and $o$ surface buoys we denote their 3D position as $\boldsymbol{S}=\{\boldsymbol{s_1},\boldsymbol{s_2},\boldsymbol{s_3},...,\boldsymbol{s_m}\}$, $\boldsymbol{R}=\{\boldsymbol{r_1},\boldsymbol{r_2},\boldsymbol{r_3},...,\boldsymbol{r_n}\}$, and $\boldsymbol{B}=\{\boldsymbol{b_1},\boldsymbol{b_2},\boldsymbol{b_3},...,\boldsymbol{b_o}\}$, respectively. Combination of the 3D positions of seabed sensors, relay sensors, and surface buoys yields $\boldsymbol{P} = \{\boldsymbol{p}_1,\boldsymbol{p}_2, \boldsymbol{p}_3,...,\boldsymbol{p}_{N}\}$ where $N=m+n+o$, $\boldsymbol{p_i}=\{x_i,y_i,z_i\}$ is the three dimension position of node $i$, and the dimensions of matrix $\boldsymbol{P}$ is ${N\times 3}$. If surface buoys, who act as anchors, are located at the same plane (i.e., located at the same depth), the 3D localization problem cannot be resolved regardless of how many surface buoys are deployed. Therefore, it is of great interest to find the optimal depth of the anchors which provide the better localization accuracy. 
\subsection{Optical Ranging for UOWSNs}
Every localization technique requires to estimate the distance between the sensor nodes or between the sensor nodes and anchors. The distance can be estimated by using different ranging techniques such as ToA, time difference of arrival (TDoA), angle of arrival (AoA), and RSS. Each of the ranging techniques has its own advantages and disadvantages, e.g., time and angle based ranging methods are more accurate as compared to the  RSS-based methods but are more complex and require extra hardware. Whereas, the RSS based methods are simple to implement but provide coarse localization. In this paper, we consider the RSS-based method where the distance between any two neighboring nodes is estimated by using the underwater optical wireless channel model. The underwater aquatic medium consists of different suspended and dissolved elements with different concentrations \cite{Haltrin:99}. These elements cause the propagation of light to suffer from absorption and scattering. Haltran's model is one of the well-known models to account for the absorption and scattering by introducing the extinction coefficient $e(\lambda)$ given as 
\begin{equation}
e(\lambda) = b(\lambda) + s(\lambda),
\end{equation}
where $b(\lambda)$ is the absorption coefficient, $s(\lambda)$ is the scattering coefficient, and $\lambda$ is the operating wavelength. The major source for the absorption of optical light in underwater aquatic medium is chlorophyll, therefore $b(\lambda)$ is expressed as
\begin{equation}
b(\lambda) = b_{wa}(\lambda)+b_{ca}(\lambda)+b_{fa} c_{fa} \exp^{-\alpha_{f}\lambda}+b_{ha} c_{ha} \exp^{-\alpha_{h}\lambda},
\end{equation}
where $\alpha_{f}$ and $\alpha_{h}$ are constants, $ b_{wa}(\lambda)$ represents the absorption in pure water, $b_{ca}(\lambda)$ represents the absorption coefficient for chlorophyll, $b_{fa} = 35.959~m^2/mg$ is the absorption coefficient of fulvic acid, $b_{ha} = 18.828~m^2/mg$ is the absorption coefficient of humic acid, $c_{fa}$ is the concentrations of fulvic acid, and $c_{ha}$ represents the concentrations of humic acid. $c_{fa}$ and $c_{ha}$ are given in \cite{Haltrin:99} as 
\begin{equation}
c_{fa} = 1.74098 c_e \exp^{(0.12327\frac{c_e}{c_o})},
\end{equation}
and
\begin{equation}
c_{ha} = 0.19334 c_e \exp^{(0.12343\frac{c_e}{c_o})},
\end{equation}
where $0 \leq c_e \leq 12~mg/m^2$ and $c_o = 1 ~ mg/m^3$. Similarly the scattering coefficient $s(\lambda)$ is modeled as
\begin{equation} \label{eq: scattering}
s(\lambda) = s_{w s} + s_{ss}(\lambda)c_{ss} + s_{sl}(\lambda)c_{ls},
\end{equation}
where {\small{$s_{ws} = 0.005826 \left(\frac{400}{\lambda}\right)^{4.322}$}} is the scattering coefficient for pure water, {\small{$s_{ss}(\lambda) = 1.151302 \left(\frac{400}{\lambda}\right)^{1.7}$}} represents the scattering coefficient for small particles, {\small{$s_{sl}(\lambda)=0.341074 \left(\frac{400}{\lambda}\right)^{0.3}$}} is the scattering coefficient for large particles, {\small{$c_{ss}=0.01739 c_e \exp{\left\{0.11631\frac{c_e}{c_o}\right\}}$}} is the concentration of small particles, and {\small{$c_{ls}=c_e \exp{\left\{0.03092\frac{c_e}{c_o}\right\}}$}} represents the concentration of large particles \cite{Haltrin:99}.

\begin{figure}
\centering
\includegraphics[width=1\columnwidth]{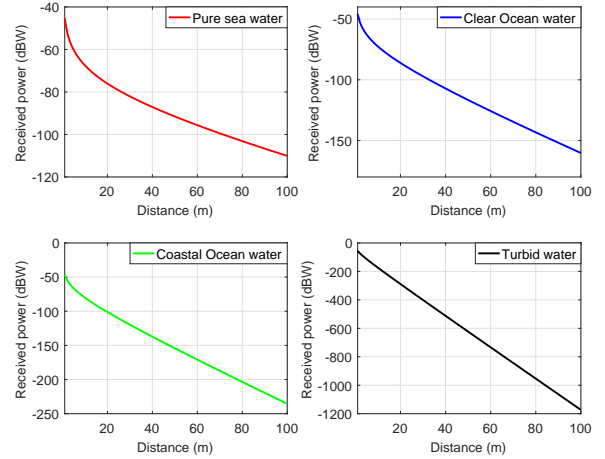}  
\caption{Received power (in dBW) Vs. distance (m) for different types of water with transmitted power of -10 dBW.\label{fig:watertype}} 
\vspace{-1.5 em} 
\end{figure}

Based on the extinction coefficient and the hardware specifications, the received power at node $j$ from node $i$ is modeled in \cite{Gkoura2017} as follows
 \begin{equation}
\label{eq:Pr}
P_{r_{i,j}} = P_{t_i} \varrho_{t_i} \varrho_{r_j} \exp^{\left(\frac{-e(\lambda) d_{ij}}{\cos \theta_{ij}}\right)}\frac{B_{r_j} \cos \theta_{ij}}{2\pi d_{ij}^2(1- \cos \theta_{0_i})}
\end{equation}
where $P_{t_i}$ is the transmission power of node $i$, $\varrho_{t_i}$ and $\varrho_{r_j}$ are the optical efficiencies of transmitter and receiver respectively, $d_{ij}$ is the distance between nodes $i$ and $j$, $\theta_{0_i}$ is the divergence angle of the transmitter, $B_{r_j}$ is the receiver aperture area, and $\theta_{ij}$ is the trajectory angle between node $i$ and $j$.  Solving \eqref{eq:Pr} for distance estimation yields
\begin{equation}\label{eq: estimateddistance}
\hat{d}_{ij} = \frac{2}{e(\lambda)} W_0 \left(\frac{e(\lambda)}{2}\sqrt{\frac{ P_{t_i}\eta_{i}\eta_j A_j\cos\theta}{P_{r_j} 2\pi (1-\cos\theta_0) }}\right)+ n_{ij}
\end{equation}
where $W_0(.)$ is the real part of Lambart $W$ function \cite{Corless1996} and $n_{ij}$ is the ranging estimation noise modeled as zero mean Gaussian random variable $n_{ij} \sim \mathcal{N}\left(0,\sigma_{ij}^2 \right)$ with variance $\sigma_{ij}^2$. Fig. \ref{fig:watertype} shows the received power as a function of distance for different types of water where the simulation parameters  are taken from  \cite{Arnon:09}. It is clear from Fig. \ref{fig:watertype} that increase in the turbidity of water reduces the received power significantly which can cause outliers in distance estimation.
In Fig. \ref{fig:watertype} only  absorption, scattering and geometrical losses are considered. However, the UOWC is also effected by other phenomena such as turbulence \cite{Oubei2017}, air bubbles \cite{JamaliKTCSAFBS16}, and salinity \cite{jamali2018}. These degrading phenomena can also introduce outliers into the distance estimation between any two nodes where the estimated distance strongly deviates from its true value. Therefore, $\hat{d}_{ij} = d_{ij}(\boldsymbol{P})+n_{ij}+o_{ij}$, where $d_{ij}(\boldsymbol{P}) =\sqrt{(x_i-x_j)^2+(y_i-y_j)^2+(z_i-z_j)^2}$ is the Euclidean distance and $o_{ij}$ is the outlier  between any two arbitrary nodes  $i$ and $j$. Including the outliers into the ranging error deviates its distribution from normal to heavy-tailed where the outliers are far away from the mean value \cite{Bryson2014}. 

\subsection{Construction of Pairwise Distance Matrix}
\begin{figure}
\centering
\includegraphics[width=1\columnwidth]{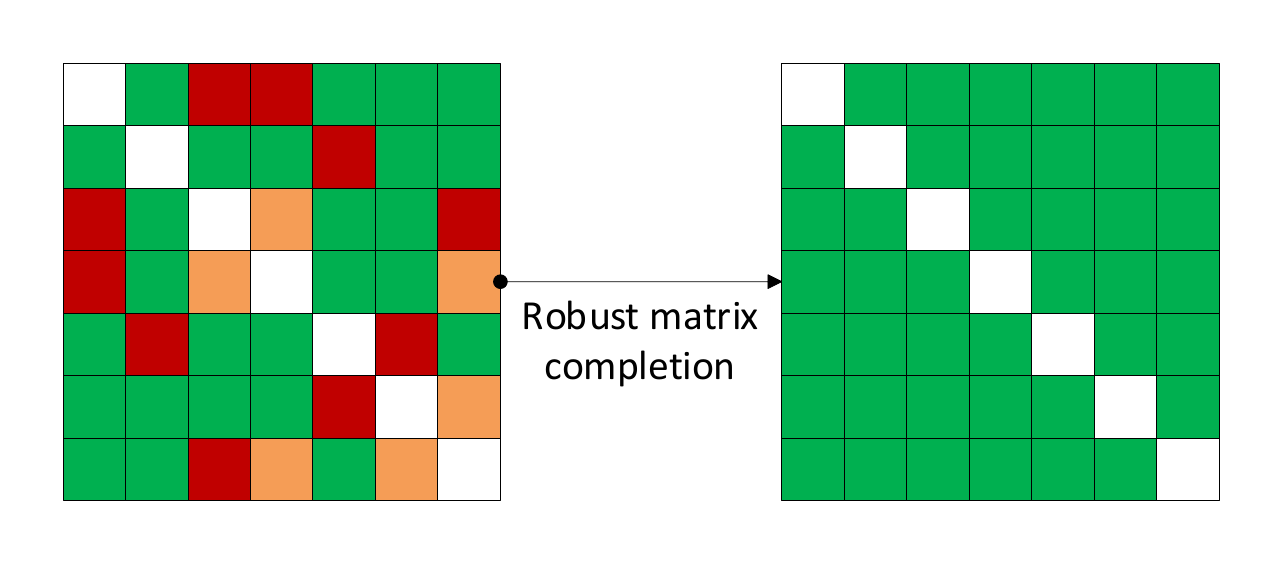}  
\caption{Matrix completion stratedy. On the left, matrix $\boldsymbol{\hat{D}}$ consists of available pairwise distances (green entries), missing distances (red entries), and outliers (orange entries). On the right side, matrix $\boldsymbol{\hat{D}}$ is completed and cleaned of the outiers.  \label{fig:snl}} 
\vspace{-1.5em} 
\end{figure}

The corresponding matrix of measured pairwise distances is denoted as $\boldsymbol{\hat{D}} = \{\hat{d}_{i,j}\}_{i=1, i\neq j}^{N}$. The noisy single hop pairwise estimated distances are collected at the surface station to form the observation distance matrix which is represented as
\begin{equation}\label{eq: distance}
\boldsymbol{\hat{D}} = \begin{bmatrix}
0 & \hat{d}_{12} & \hat{d}_{13} & \cdots & \hat{d}_{1N}\\
\hat{d}_{21} & 0 & \hat{d}_{23} & \cdots & \hat{d}_{2N}\\
\hat{d}_{31} & \hat{d}_{32} & 0 & \cdots & \hat{d}_{3N}\\
\vdots & \vdots & \vdots & \ddots & \vdots\\
\hat{d}_{N1} & \hat{d}_{N2} & \hat{d}_{N3}& \cdots & 0
\end{bmatrix}.
\end{equation}
However, it is very rare to directly obtain all the pairwise distances because of harsh propagation characteristics of UOWC channels. Instead, a subset of pairwise distances are available which are noisy and there is a great possibility to get outliers. Accordingly, matrix $\boldsymbol{\hat{D}}$ can be re-written as
\begin{equation}\label{eq: distance2}
\boldsymbol{\hat{D}} = \begin{bmatrix}
0 & \hat{d}_{12} & ? & \cdots & \hat{d}_{1N}\\
\hat{d}_{21} & 0 & {o}_{23} & \cdots & \hat{d}_{2N}\\
\hat{d}_{31} & {o}_{32} & 0 & \cdots & {o}_{3N}\\
\vdots & \vdots & \vdots & \ddots & \vdots\\
\hat{d}_{N1} & ? & o_{3N}& \cdots & 0
\end{bmatrix},
\end{equation}
where $?$ are the missing pairwise distances and $o_{ij}$ represents the outliers. Hence, goal of the proposed localization method is to determine the 3D positions of seabed sensors and relay sensors, i.e., $\boldsymbol{\tilde{P}} = \{\boldsymbol{\tilde{p}}_1,\boldsymbol{\tilde{p}}_2, \boldsymbol{\tilde{p}}_3,...,\boldsymbol{\tilde{p}}_{m+n}\}$ from  matrix $\boldsymbol{\hat{D}}$. As mentioned before that matrix $\boldsymbol{\hat{D}}$  has missing distances as well as outliers, therefore, it is first required to complete the missing distances and remove the outliers in $\boldsymbol{\hat{D}}$ as illustrated in Fig.~\ref{fig:snl}.
\section{Proposed 3D Localization Method}\label{sec:proposed}
In this section, first, we propose a novel low-rank matrix completion and outliers removal strategy for 3D localization of UOWSNs with limited connectivity. Matrix completion assigns values to the missing entries of a partially filled matrix and removes the outliers. Secondly, we investigate the impact of anchors placement on the localization accuracy and optimize the placement (especially the depth) of anchors to minimize the average localization error.

\subsection{Low Rank Matrix Completion and Outliers Removal}
A number of conventional matrix completion methods have been proposed in the past, for example, singular value threshold \cite{Cai2010}, Optspace \cite{Keshavan2010}, augmented Lagrange multiplier method \cite{Lin2013, Wang2016}, Poisson matrix completion \cite{Cao2015}, and alternating minimization \cite{Gamarnik2016}. All of aforementioned matrix completion methods are not robust to the outliers present in matrix $\boldsymbol{\hat{D}}$ and consider a least square loss function which measures the raw stress between $\hat{d}_{ij}$ and $d_{ij}$, i.e.,
\begin{equation}\label{stress}
\sigma(\boldsymbol{P}) = \sum_{j>i}^N w_{ij}\left(\hat{d}_{ij}-d_{ij}(\boldsymbol{P})\right)^2,
\end{equation}
where $w_{ij}$ represents the user-defined nonnegative weights to give importance to the measured distances. In majority of the cases, weights are considered to be equal to one. However, there are several other cases where unequal weights are employed such as Sammon mapping \cite{Sammon1969} with $w_{ij}= (\hat{d}_{ij})^{-1}$ and elastic scaling \cite{BMSP367} with $w_{ij} = (\hat{d}_{ij})^{-2}$. 
Error propagation for each element of matrix $\boldsymbol{\hat{D}}$ during the double centering method is given by $-\frac{1}{2} \boldsymbol{C}\boldsymbol{\hat{D}}^2 \boldsymbol{C}$  where $\boldsymbol{C} = \boldsymbol{I} - \frac{\boldsymbol{1}\boldsymbol{1}^T}{N}$ is known as the centering operation, $\boldsymbol{I}$ is  an $N \times N$ identity matrix, $\boldsymbol{1}$ is the $N \times 1$ vector of ones, and $T$ is the transpose operator \cite{Andreas2008}. Following from the error propagation of the double centering method, it is obvious that even a single outlier in $\boldsymbol{\hat{D}}$ can severely distort the solution for \eqref{stress}. Here the outliers are referred to the measured distances which are significantly different than the actual Euclidean distances.  Therefore, it is necessary to investigate methods which can remove these outliers. Even if the problem defined in \eqref{stress} is a non-convex optimization problem  without a unique solution, it can be solved by iterative majorization approach which minimizes  
\begin{equation}
\sigma(\boldsymbol{P}) = \parallel -\frac{1}{2}\boldsymbol{C}[\boldsymbol{\hat{D}}^2 - \boldsymbol{D}^2]\boldsymbol{C}\parallel_F^2,
\end{equation}
where $\parallel \cdot \parallel_F$ is the Frobenius norm.  The effect of double centering method can easily be compensated by modifying the error function to $l_1$ norm, i.e., $\parallel \boldsymbol{\hat{D}}^2 - \boldsymbol{D}^2 \parallel_1$. However, $l_1$ norm is not smooth due to the fact that it has a singularity at its origin. To mitigate this problem, the use of Huber's loss function can be of great benefit which interpolates between $l_1$ and $l_2$ norms minimizations, i.e., 
\begin{equation}\label{eq: huber1}
\sigma_h(\boldsymbol{P}) = \sum_{j>i}^N w_{ij} \gamma(a)\left(\hat{d}_{ij}-d_{ij}(\boldsymbol{P})\right)^2,
\end{equation}
where $\gamma(a)$ is the Huber's loss function given as,
\begin{equation}
\gamma(a)=\begin{cases}
\frac{a^2}{2}~~~~~~~~~~~\text{if}~|a|\leq \rho \\
\rho |a|-\frac{\rho^2}{2}~~~\text{if}~|a| > \rho
\end{cases},
\end{equation}
 $a$ is the argument of the Huber's loss function which represents the residual error and $\rho$ is the threshold which can be chosen adaptively or arbitrarily from matrix $\boldsymbol{\hat{D}}^2$. It is argued by Huber that if  $\rho = 1.345$ then Huber loss function is  95\% as efficient as the least square solution \cite{Udell2016}. Similarly, the loss function proposed by Tukey provide the same result with $\rho = 4.685$ where 
\begin{equation}
\gamma(a)=\begin{cases}
a\left(1-(\frac{a}{\rho})^2\right)^2~~~~~~~~~~~\text{if}~|a|\leq \rho \\
0~~~~~~~~~~~~~~~~~~~~~~~~~~~\text{if}~|a| > \rho
\end{cases}.
\end{equation}
As the estimated distance between any two nodes $i$ and $j$ is modeled as $\hat{d}_{ij} = d_{ij}+n_{ij}+o_{ij}$, where $o_{ij}$ represents the outliers and $n_{ij}$ is a zero mean random variable to model the nominal errors. In the presence of outliers and nominal noise error, the problem defined in \eqref{eq: huber1} follows $n$-contaminated distribution and therefore adopts Huber's function for the residuals \cite{Huber2009}. Nonetheless, the outliers are less in number and sparse, therefore inserting the $l_1$ norm minimization problem into \eqref{stress} is justified and yields
\begin{eqnarray}\label{eq: minimization}
(\hat{\boldsymbol{P}}, \hat{\boldsymbol{O}})&=&\argmin_{{\boldsymbol{P}}, {\boldsymbol{O}}}\left\{\sum_{i<j}^N w_{ij}\left(\hat{d}_{ij}-d_{ij}(\boldsymbol{P})- o_{ij}\right)^2 \right. \nonumber \\
& & \left. +\lambda_1\sum_{j>i}^N|o_{ij}|\right\},
\end{eqnarray} 
where matrix $\boldsymbol{\hat{O}}$ represents the estimated outliers.  In order to get to the solution in \eqref{eq: minimization}, the following function needs to be minimized
\begin{eqnarray}\label{eq: minimization2}
\Theta({\boldsymbol{P}}, {\boldsymbol{O}})&=&\min_{{\boldsymbol{P}}, {\boldsymbol{O}}}\left\{\sum_{j>i}^N w_{ij}\left(\hat{d}_{ij}-d_{ij}(\boldsymbol{P})- o_{ij}\right)^2 \right. \nonumber \\
& & \left. +\lambda_1\sum_{i>j}^N|o_{ij}|\right\}.
\end{eqnarray} 
The first term in \eqref{eq: minimization2} corresponds to the level of fitness between $\hat{d}_{ij}$ and $d_{ij}$ after removing the outlier $o_{ij}$ and the second term corresponds to the penalty linked to the sparsity of matrix $\boldsymbol{O}$ where $\lambda_1$ represents the regularization parameter. The function in \eqref{eq: minimization2} is non-convex and non-differentiable therefore expanding it yields
\begin{eqnarray}\label{eq: minimization3}
\Theta({\boldsymbol{P}}, {\boldsymbol{O}})&=&\min_{{\boldsymbol{P}}, {\boldsymbol{O}}}\left\{\frac{1}{2}\parallel \hat{\boldsymbol{D}} - \boldsymbol{D} \parallel_F^2 \right. \nonumber \\ & & -2 \sum_{j>i}(\hat{d}_{ij}-o_{ij})d_{ij}(\boldsymbol{P}) \left. + \sum_{j>i}d_{ij}^2(\boldsymbol{P})\right. \nonumber \\ & &+ \left. \frac{\lambda_1}{2}\parallel \boldsymbol{O} \parallel_1 \right\}.
\end{eqnarray} 
where $\parallel\boldsymbol{O}\parallel_1 = \sum_{j>i}|o_{ij}|$. Consider that $\{\boldsymbol{v}_n \in \mathrm{R}^N\}_{n=1}^N$ is a set of indicator vectors with elements $[\boldsymbol{v}_i]_j = 1$ if $i = j$ and zero if $i \neq j$. Then the term $\sum_{j>i}d_{ij}^2(\boldsymbol{P})$ in \eqref{eq: minimization3} can be written as
\begin{eqnarray}\label{eq: d1}
\sum_{j>i}d_{ij}^2(\boldsymbol{P}) &=& \sum_{j>i} (\boldsymbol{v}_i - \boldsymbol{v}_j)^T\boldsymbol{P}^T \boldsymbol{P} (\boldsymbol{v}_i - \boldsymbol{v}_j), \nonumber \\
& & = \text{Tr}(\boldsymbol{P}\boldsymbol{\mathcal{L}}\boldsymbol{P}^T),
\end{eqnarray}
where $\text{Tr}(\cdot)$ is the trace operator and ${\boldsymbol{\boldsymbol{\mathcal{L}}}}$ is a $N \times N$  matrix with non-diagonal elements equal to $-1$ and diagonal elements equal to $N-1$. The term $-2 \sum_{j>i}(\hat{d}_{ij}-o_{ij})d_{ij}(\boldsymbol{P})$ in \eqref{eq: minimization3} is non-convex and non-differentiable especially when $\hat{d}_{ij} > o_{ij}$. Majorization of        this term is required to minimize \eqref{eq: minimization3}, therefore, defining a majorization function $m({\boldsymbol{P}}, {\boldsymbol{O}}) = -2 \sum_{j>i}(\hat{d}_{ij}-o_{ij})d_{ij}(\boldsymbol{P})$. $m({\boldsymbol{P},\boldsymbol{O}})$ can be split into two terms depending on the value of $\hat{d}_{ij}-o_{ij}$, i.e., $ m({\boldsymbol{P}}, {\boldsymbol{O}})= m_{+}({\boldsymbol{P}}, {\boldsymbol{O}})+m_{-}({\boldsymbol{P}}, {\boldsymbol{O}})$, where
\begin{equation}
m_{+}({\boldsymbol{P}}, {\boldsymbol{O}}) = -2 \sum_{j>i}(\hat{d}_{ij}-o_{ij})d_{ij}(\boldsymbol{P}),~~\text{for}~\hat{d}_{ij} > o_{ij},
\end{equation}
and
\begin{equation}
m_{-}({\boldsymbol{P}}, {\boldsymbol{O}}) = -2 \sum_{j>i}(\hat{d}_{ij}-o_{ij})d_{ij}(\boldsymbol{P}),~~\text{for}~\hat{d}_{ij} < o_{ij}.
\end{equation}
Majorization of $m_{+}({\boldsymbol{P}}, {\boldsymbol{O}})$ is obtained by using Cauchy-Schwartz inequality, given as
\begin{equation}\label{eq: m1}
m_{+}({\boldsymbol{P}}, {\boldsymbol{O}}) \leq -2 \sum_{j>i}\left(\frac{\hat{d}_{ij}-o_{ij}}{d_{ij}({\boldsymbol{P}})}\right)(\boldsymbol{z}_i - \boldsymbol{z}_j)^T(\boldsymbol{p}_i - \boldsymbol{p}_j),
\end{equation}
where $\boldsymbol{Z} = \{\boldsymbol{z}_1,\boldsymbol{z}_2,...,\boldsymbol{z}_N\} \in \mathcal{R}^{3\times N}$ is a matrix of auxiliary variables. Majorization of $m_{-}({\boldsymbol{P}}, {\boldsymbol{O}})$ is obtained by using the fact that $\sqrt{y}$ is upper bounded by $\sqrt{y} \leq \sqrt{y_0}+ \frac{1}{\sqrt{y_0}}(y-y_0)$ for any positive value of $y_0$. Consider that $y = d_{ij}^2(\boldsymbol{P})$ and $y_0 = d_{ij}^2(\boldsymbol{Z})$, then using the bound leads to
\begin{equation}\label{eq: m2}
m_{-}({\boldsymbol{P}}, {\boldsymbol{O}}) \leq -2 \sum_{j>i}\left(\frac{\hat{d}_{ij}-o_{ij}}{d_{ij}({\boldsymbol{P}})}\right) d_{ij}^2(\boldsymbol{P}) + B_0({\boldsymbol{Z}}, {\boldsymbol{O}}),
\end{equation}
where $B_0({\boldsymbol{Z}}, {\boldsymbol{O}}) = - \sum_{j>i} (\hat{d}_{ij}-o_{ij})d_{ij}(\boldsymbol{P})$ for $\hat{d}_{ij} > o_{ij}$. Finally adding \eqref{eq: m1} and \eqref{eq: m2} yields
\begin{eqnarray}\label{eq: m3}
m({\boldsymbol{P}}, {\boldsymbol{Z}},{\boldsymbol{O}}) &=& -2 \text{Tr}\left(\boldsymbol{P}\boldsymbol{L}_+(\boldsymbol{Z},\boldsymbol{O})\boldsymbol{Z}^T\right) \\
& & +\text{Tr}\left(\boldsymbol{P}\boldsymbol{L}_+(\boldsymbol{Z},\boldsymbol{O})\boldsymbol{P}^T\right)+B_0({\boldsymbol{Z}}, {\boldsymbol{O}}),\nonumber
\end{eqnarray}
where $\boldsymbol{\boldsymbol{\mathcal{L}}}_+(\boldsymbol{Z},\boldsymbol{O})$ is $N \times N$ Laplacian matrix. When equality holds between $\boldsymbol{Z}$ and $\boldsymbol{P}$, then elements of the Laplacian matrix are defined as \cite{Aflalo2013}
\begin{equation}
\boldsymbol{\boldsymbol{\mathcal{L}}}_+(\boldsymbol{P},\boldsymbol{O})_{ij} = \begin{cases}
-(\hat{d}_{ij}-o_{ij})d_{ij}^{-1}(\boldsymbol{P})~~~~~~(i,j)\in \mathbb{A}\\
~0~~~~~~~~~~~~~~~~~~~~~~~~~~~~(i,j)\in \mathbb{B}\\
-\sum_{k=1,k \neq i}^N\boldsymbol{\boldsymbol{\mathcal{L}}}_+(\boldsymbol{O},\boldsymbol{P})_{ik}~ (i,j)\in \mathbb{C}
\end{cases}
\end{equation}
where $\mathbb{A}(\boldsymbol{P},\boldsymbol{O})=\{(i,j):i\neq j,~ d_{ij}(\boldsymbol{P})\neq 0,~ \hat{d}_{ij}> o_{ij})\}$, $\mathbb{B}(\boldsymbol{P},\boldsymbol{O})=\{(i,j):i\neq j,~ d_{ij}(\boldsymbol{P})= 0,~ \hat{d}_{ij}> o_{ij})\}$, and $\mathbb{C}(\boldsymbol{P},\boldsymbol{O})=\{(i,j):i=j,~ \hat{d}_{ij}> o_{ij})\}$.
Substituting \eqref{eq: m3} and \eqref{eq: d1} in \eqref{eq: minimization3} yields 
\begin{eqnarray}\label{eq: minimization4}
\Theta({\boldsymbol{P}}, {\boldsymbol{O}})&=&  \text{Tr}(\boldsymbol{P}\boldsymbol{\mathcal{L}}\boldsymbol{P}^T)- \text{Tr}\left(\boldsymbol{P}\boldsymbol{L}_+(\boldsymbol{P},\boldsymbol{O})\boldsymbol{P}^T\right) \\
& & +\frac{1}{2}\parallel \hat{\boldsymbol{D}} - \boldsymbol{D} \parallel_F^2 +\frac{\lambda_1}{2}\parallel \boldsymbol{O} \parallel_1 +B_0({\boldsymbol{Z}}, {\boldsymbol{O}}).\nonumber
\end{eqnarray} 
The majorization function $\Theta(\cdot)$ in \eqref{eq: minimization4} is convex and differentiable in terms of both $\boldsymbol{O}$ and $ \boldsymbol{P}$. Therefore, solving \eqref{eq: minimization4} by minimization, i.e.,
\begin{equation}\label{eq: outr2}
\boldsymbol{O}^{k+1} = \argmin_{\boldsymbol{O}}\mathrm{s}({\boldsymbol{P}^k},{\boldsymbol{O}}).
\end{equation}
Indeed, each entry $o_{ij}^{k+1}$ of \eqref{eq: outr2} corresponds to a standard least absolute shrinkage and selection operator (LASSO) solution which is expressed as
\begin{equation}\label{eq: out1}
o_{ij}^{k+1} = \mathcal{S}_{\lambda_{1}}(\hat{d}_{ij}-d_{ij}(\boldsymbol{P}^{k})),
\end{equation}
where $\mathcal{S}_{\lambda_{1}}(x) = \text{sign}(x)(|x|-\frac{\lambda_1}{2})_{+}$ is the soft thresholding operator, and $(\cdot )_+  = \max (0,\cdot)$ \cite{Robert1996}. Similarly, the position estimates are obtained by minimizing the following function
\begin{equation}\label{eq: out2}
\boldsymbol{P}^{k+1} = \argmin_{\boldsymbol{P}}\mathrm{s}( {\boldsymbol{P}^k},{\boldsymbol{O}^{k+1}}).
\end{equation}
The closed-form solution of \eqref{eq: out2} is obtained by using the first order optimality condition \cite{ivan2012}, i.e.,
\begin{equation}\label{eq: leastout}
\boldsymbol{P}^{k+1}= {\boldsymbol{\boldsymbol{\mathcal{L}}}^\dagger}\boldsymbol{\boldsymbol{\mathcal{L}}}_+(\boldsymbol{O}^{k+1},\boldsymbol{P}^{k})\boldsymbol{P}^{k},
\end{equation}
where ${\boldsymbol{\boldsymbol{\mathcal{L}}}^\dagger}=\boldsymbol{C}/N$ is a Moore-Penrose pseudo-inverse of matrix ${\boldsymbol{\boldsymbol{\mathcal{L}}}}$.  For this iterative process, the initial configuration of $\boldsymbol{P}^{0}$ is randomly chosen whereas the elements of the outlier matrix $\boldsymbol{O}^{0}$ are set to zero. Given $\boldsymbol{P}^{k}$, the estimation of $\boldsymbol{O}^{k+1}$  via \eqref{eq: out1} is a $l_1$ regularization problem while for a given $\boldsymbol{O}^{k+1}$ the estimation of $\boldsymbol{P}^{k+1}$ is the least square optimization problem, i.e., $\parallel \boldsymbol{\boldsymbol{\mathcal{L}}} \boldsymbol{P}^{k+1} - \boldsymbol{\boldsymbol{\mathcal{L}}}_+(\boldsymbol{O}^{k+1},\boldsymbol{P}^{k})\boldsymbol{P}^{k} \parallel_F^2$. Even though  the estimation of $\boldsymbol{P}^{k+1}$ accounts for the outliers, it is still a least square solution and is greatly influenced by the outliers. Therefore,  it is proposed to apply function $f(\cdot)$ on to the residual $\boldsymbol{\boldsymbol{\mathcal{L}}} \boldsymbol{P}^{k+1} - \boldsymbol{\boldsymbol{\mathcal{L}}}_+(\boldsymbol{O}^{k+1},\boldsymbol{P}^{k})\boldsymbol{P}^{k}$  in \eqref{eq: leastout} where $f(\cdot)$ is non-negative and differentiable with respect to $\boldsymbol{P}$ to impose smoothness such that
\begin{eqnarray}\label{smooth}
\boldsymbol{P}^{k+1} & = & \argmin_{\boldsymbol{P}}\left( f(\boldsymbol{\boldsymbol{\mathcal{L}}} \boldsymbol{P} - \boldsymbol{\boldsymbol{\mathcal{L}}}_+(\boldsymbol{O}^{k+1},\boldsymbol{P}^{k})\boldsymbol{P}^{k})\right.\nonumber \\ & & \left. +\lambda_2\parallel\boldsymbol{P}\parallel_F^2\right).
\end{eqnarray}
The optimization problem defined in \eqref{smooth} can be solved by using half quadratic minimization (HQM).  HQM was pioneered to reconstruct the signal and images from  corrupted signal and images \cite{golub2012matrix}. Recently, HQM functions have been widely used in computer vision, machine learning, and feature extraction \cite{zhang2009multi}. Based on the theory of the conjugate function in HQM, the problem in \eqref{smooth} can be written as \cite{golub2012matrix}
\begin{equation}\label{eq: hqminimize}
\Phi(\tilde{\boldsymbol{\boldsymbol{\mathcal{L}}}}^{k},\boldsymbol{A})=\min_{\boldsymbol{A}}\left(Q(\tilde{\boldsymbol{\boldsymbol{\mathcal{L}}}}^{k}, \boldsymbol{A})+\sum_{i=1}^N \sum_{j=1}^3\psi(a_{i,j})\right),
\end{equation}
where $\Phi(\cdot )$ is known as potential loss function, $\tilde{\boldsymbol{\boldsymbol{\mathcal{L}}}}^{k} = \boldsymbol{\boldsymbol{\mathcal{L}}} \boldsymbol{P} - \boldsymbol{\boldsymbol{\mathcal{L}}}_+(\boldsymbol{O}^{k+1},\boldsymbol{P}^{k})\boldsymbol{P}^{k}$, $\boldsymbol{A}$ is $N \times 3$ matrix of auxiliary variables, $Q(\cdot )$ is the quadratic function, and $\psi(\cdot)$ is the conjugate function of $\Phi(\cdot )$. The quadratic function in \eqref{eq: hqminimize} is given in \cite{golub2012matrix} as
\begin{equation}\label{eq: qfunction}
Q(\tilde{\boldsymbol{\boldsymbol{\mathcal{L}}}}^{k}, \boldsymbol{A})=\parallel \sqrt{c}~(\tilde{\boldsymbol{\boldsymbol{\mathcal{L}}}}^{k})-\frac{\boldsymbol{A}}{\sqrt{c}} \parallel_F^2,
\end{equation}
where $c$ is a constant parameter. Substituting \eqref{eq: qfunction} into \eqref{eq: hqminimize} and adding the regularization parameter for $\boldsymbol{P}$ yields 
\begin{eqnarray}\label{eq: altminimize}
\Phi(\boldsymbol{P}, \boldsymbol{A})&=& \min_{\boldsymbol{P},\boldsymbol{A}}\bigg(\parallel\sqrt{c}~(\tilde{\boldsymbol{\boldsymbol{\mathcal{L}}}}^{k})-\frac{\boldsymbol{A}}{\sqrt{c}} \parallel_F^2+\sum_{i=1}^N \sum_{j=1}^3\psi(a_{i,j})\bigg. \nonumber \\ & & \bigg.
+ \lambda_2\parallel\boldsymbol{P}\parallel_F^2\bigg).
\end{eqnarray} 
 The minimization function in \eqref{eq: altminimize}  is called resultant cost function of $\Phi(\cdot)$, and is solved by using alternating minimization given in \cite{He2014} as 
\begin{equation}
\boldsymbol{A}^{k+1} = \mu(\boldsymbol{\boldsymbol{\mathcal{L}}}\boldsymbol{P}^{k}-\boldsymbol{\boldsymbol{\mathcal{L}}}_+(\boldsymbol{O}^{k+1},\boldsymbol{P}^{k})\boldsymbol{P}^{k})
\end{equation}
where $\mu(\cdot)$ is the minimizer function for auxiliary variables and 
\begin{equation}\label{eq: altminimize2}
\boldsymbol{P}^{k+1} = \argmin_{\boldsymbol{P}}\left(\parallel \sqrt{c}~(\tilde{\boldsymbol{\boldsymbol{\mathcal{L}}}}^{k})-\frac{\boldsymbol{A}^{k+1}}{\sqrt{c}} \parallel_F^2 + \lambda_2\parallel\boldsymbol{P}\parallel_F^2\right).
\end{equation}
The minimization problem in \eqref{eq: altminimize2} can be re-written as
\begin{eqnarray}\label{eq: altminimize3}
\boldsymbol{P}^{k+1} & = & \argmin_{\boldsymbol{P}}\left\{c~\text{Tr}\left((\boldsymbol{\boldsymbol{\mathcal{L}}}\boldsymbol{A}-\boldsymbol{R}^{k+1})^T(\boldsymbol{\boldsymbol{\mathcal{L}}}\boldsymbol{A}-\boldsymbol{R}^{k+1}) \right)\right. \nonumber \\  & & \left. +\lambda_2 \text{Tr}(\boldsymbol{P}^T\boldsymbol{P})\right\}.
\end{eqnarray}
where $\boldsymbol{R}^{k+1} = \boldsymbol{\boldsymbol{\mathcal{L}}}_+(\boldsymbol{O}^{k+1},\boldsymbol{P}^{k})\boldsymbol{P}^{k} + \frac{\boldsymbol{A}^{k+1}}{c}$. The closed-form solution of \eqref{eq: altminimize3} with respect to $\boldsymbol{P}$ is obtained by using the first order optimality condition, i.e.,
\begin{equation}
\hat{\boldsymbol{P}}^{k+1} = c~(c \boldsymbol{\boldsymbol{\mathcal{L}}}^T \boldsymbol{\boldsymbol{\mathcal{L}}} + \lambda_2 \boldsymbol{I})^{-1}\boldsymbol{\boldsymbol{\mathcal{L}}}^T\boldsymbol{R}^{k+1},
\end{equation}
where $\hat{\boldsymbol{P}}^{k+1}$ is an $N \times 3$ matrix which constitutes of the 3D relative position estimation of each node in the network. After getting the relative position estimations $\hat{\boldsymbol{P}}$, the obtained solution can be transformed into the global position estimation $\tilde{\boldsymbol{P}}$ by using the positions of anchors in the network. The common procedure for transformation to the global position estimation is orthogonal Procrustes analysis given by \cite{Colin1991}
\begin{equation}
\tilde{\boldsymbol{P}}=\beta(\hat{\boldsymbol{P}})\boldsymbol{\Omega}+\boldsymbol{\upsilon},
\end{equation}
where $\beta$, $\boldsymbol{\Omega}$, and $\boldsymbol{\upsilon}$ are the scaling, rotation, and translation elements respectively.
\subsection{Optimal Placement of Anchors}
Optimal anchor placement is crucial for better position estimation. Therefore, in this section, we develop an optimal anchor placement strategy to improve the localization accuracy of the proposed technique.  As in the previous sections, the pairwise distances are estimated between every node in the network, now it is possible to optimize the position of anchors to reduce the localization error. Consider that there are $o$ number of surface buoys in the network which also work as  anchors, then the estimated distance between any arbitrary sensor node and $k$-th surface buoy can be written as
\begin{equation}
\hat{d}_k = d_k + n_k,~~~~~~~~~ k= 1,2,...,o,
\end{equation} 
where $d_k = \sqrt{(x-x_k)^2+(y-y_k)^2+(z-z_k)^2}$ is the Euclidean distance to $k$-th anchor and $n_k$ is the corresponding range measurement error.  The Cramer Rao lower bound (CRLB) of  the $j$-th parameter of $\boldsymbol{p}$ is defined as
\begin{equation}
\mathcal{C}(\boldsymbol{\tilde{p}})=\boldsymbol{{J}}^{-1}(\boldsymbol{p})_{j,j},
\end{equation}
where $\boldsymbol{{J}}(\boldsymbol{p})$ represents the Fisher information matrix (FIM). The elements of FIM are given as
\begin{equation}\label{fisher1}
\boldsymbol{{J}}(\boldsymbol{p})_{j,k}= -\textit{E}\left(\frac{\partial^2 \log f(\boldsymbol{\hat{d}}|\boldsymbol{p})}{\partial \boldsymbol{p}_j \partial \boldsymbol{p}_k} \right),~~~~j,k = 1,2,3,
\end{equation}\
where  \textit{E} is the expectation operator and $\boldsymbol{\hat{d}} = \{\hat{d}_1,\hat{d}_2,...,\hat{d}_o\}$ is a vector of size $o \times 1$ containing the estimated distances of a single node to all the anchors. Since the ranging measurements are effected by Gaussian noise, then the probability density function $f(\boldsymbol{\hat{d}}|\boldsymbol{p})$ for the range measurements is defined as 
\begin{equation}
f(\boldsymbol{\hat{d}}|\boldsymbol{p}) = \frac{1}{\sqrt{(2\pi)^o}\prod_{i=1}^o \sigma_i} \exp^{\left(- \sum_{i=1}^o \frac{1}{2\sigma_i^2}(\hat{d}_i-d_i)^2 \right)}.
\end{equation}
The sub-matrices of $\boldsymbol{{J}}(\boldsymbol{p})_{j,k}$ are derived from \eqref{fisher1} and given as 
\begin{equation}
\boldsymbol{{J}}(\boldsymbol{p})_{j=1,k=1} = \sum_{i=1}^o\left( \frac{(x-x_i)^2}{(\sigma_i d_i)^2}\right),
\end{equation}
\begin{equation}
\boldsymbol{{J}}(\boldsymbol{p})_{j=1,k=2} = \boldsymbol{{J}}(\boldsymbol{p})_{j=2,k=1}=\sum_{i=1}^o\left( \frac{(x-x_i)(y-y_i)}{(\sigma_i d_i)^2}\right),
\end{equation}
\begin{equation}
\boldsymbol{{J}}(\boldsymbol{p})_{j=1,k=3} = \boldsymbol{{J}}(\boldsymbol{p})_{j=3,k=1}=\sum_{i=1}^o\left( \frac{(x-x_i)(z-z_i)}{(\sigma_i d_i)^2}\right),
\end{equation}
\begin{equation}
\boldsymbol{{J}}(\boldsymbol{p})_{j=2,k=2} = \sum_{i=1}^o\left( \frac{(y-y_i)^2}{(\sigma_i d_i)^2} \right),
\end{equation}
\begin{equation}
\boldsymbol{{J}}(\boldsymbol{p})_{j=2,k=3} = \boldsymbol{{J}}(\boldsymbol{p})_{j=3,k=2}=\sum_{i=1}^o\left( \frac{(y-y_i)(z-z_i)}{(\sigma_i d_i)^2}\right),
\end{equation}
\begin{equation}
\boldsymbol{{J}}(\boldsymbol{p})_{j=3,k=3} = \sum_{i=1}^o\left( \frac{(z-z_i)^2}{(\sigma_i d_i)^2}\right).
\end{equation}
\begin{figure*}
    \centering
    \begin{subfigure}[b]{0.45\textwidth}
\includegraphics[width=1\columnwidth]{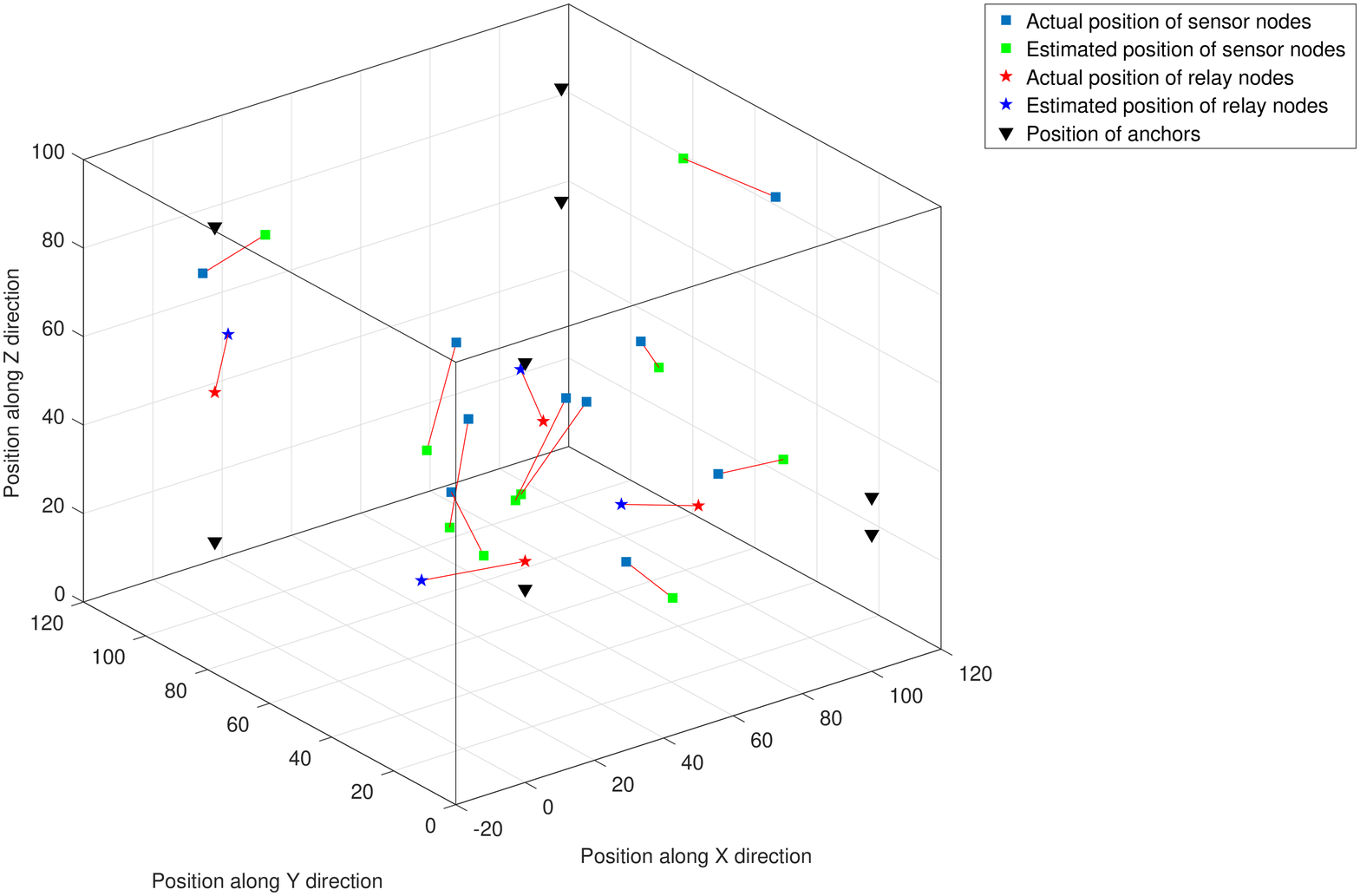}  
\caption{Localization of $m = 10$ and $n=4$ in a volume of 100 $m^3$ with random depth of anchors and in presense of outliers (RMSE = 23.70 m) .\label{fig:outlierrandom}}  
    \end{subfigure}
    \begin{subfigure}[b]{0.48\textwidth}
\includegraphics[width=1\columnwidth]{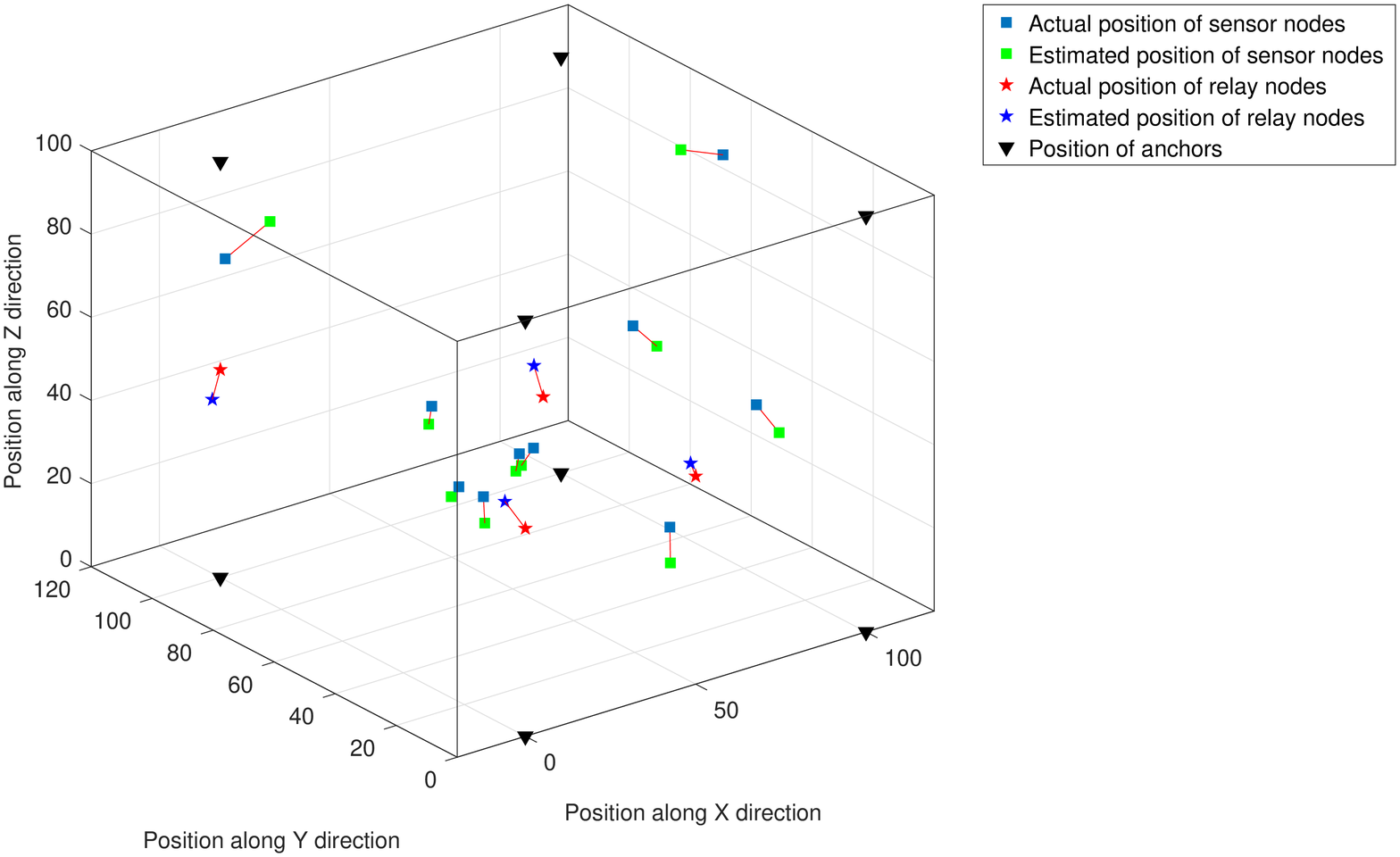}  
\caption{Localization of $m = 10$ and $n=4$ in a volume of 100 $m^3$ with optimal depth of anchors and in presense of outliers (RMSE = 8.46 m).\label{fig:outlieroptimum}}  
    \end{subfigure}
    
    \begin{subfigure}[b]{0.48\textwidth}
\includegraphics[width=1\columnwidth]{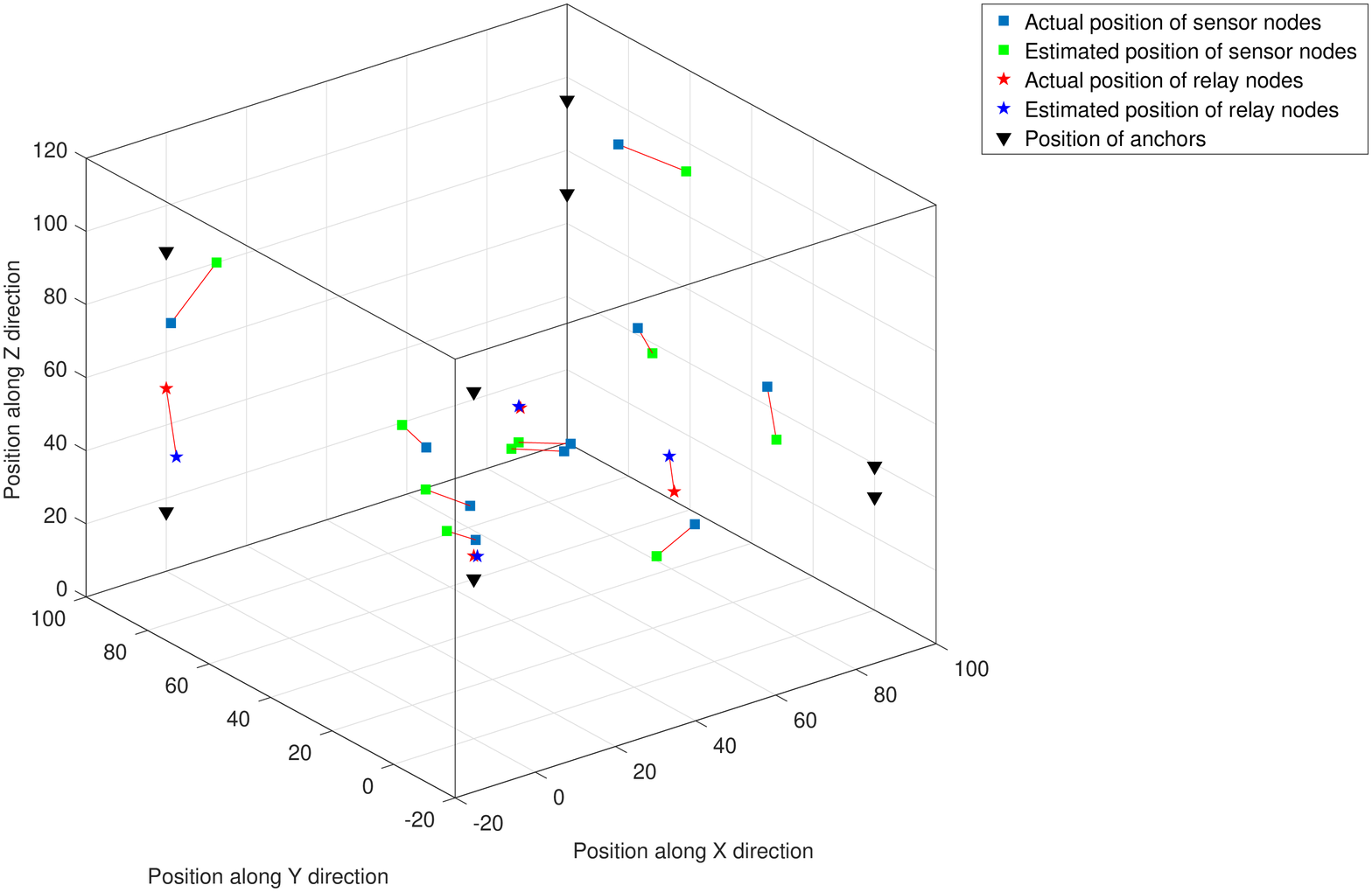}  
\caption{Localization of $m = 10$ and $n=4$ in a volume of 100 $m^3$ with removal of outliers and at random depth of anchors (RMSE = 11.64 m).\label{fig:nooutlierrandom}}  
    \end{subfigure}
    \begin{subfigure}[b]{0.48\textwidth}
\includegraphics[width=1\columnwidth]{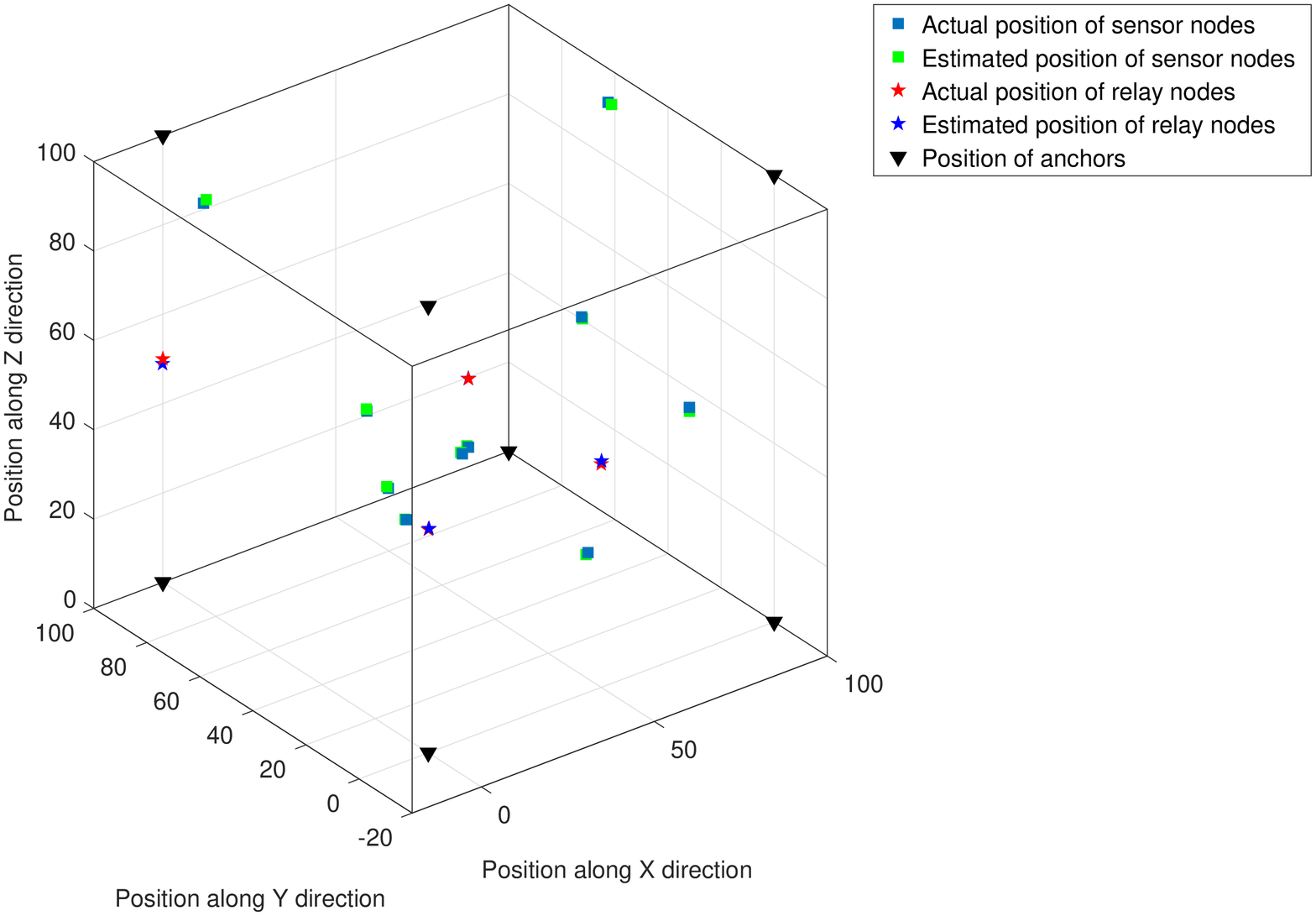}  
\caption{Localization of $m = 10$ and $n=4$ in a volume of 100 $m^3$ with removal of outliers and at random depth of anchors (RMSE = 0.659 m).\label{fig:nooutlieroptimum}}  
    \end{subfigure}    
\caption{3D localization of UOWSNs}
\label{fig:prop_localization}
\hrule
\vspace{-1.5 em}
\end{figure*}
Now the CRLB for the three-dimensional position estimation is obtained as
\begin{eqnarray}
\mathcal{C}({\tilde{x}})&=&\frac{1}{\text{det}(\boldsymbol{{J}}(\boldsymbol{p}))}\Big(\boldsymbol{{J}}(\boldsymbol{p})_{j=2,k=2}\boldsymbol{{J}}(\boldsymbol{p})_{j=3,k=3}\Big. \nonumber \\ & & \Big.-\boldsymbol{{J}}(\boldsymbol{p})_{j=2,k=3}^2\Big),
\end{eqnarray}
\begin{eqnarray}
\mathcal{C}({\tilde{y}})&=&\frac{1}{\text{det}(\boldsymbol{{J}}(\boldsymbol{p}))}\Big(\boldsymbol{{J}}(\boldsymbol{p})_{j=1,k=1}\boldsymbol{{J}}(\boldsymbol{p})_{j=3,k=3}\Big. \nonumber \\ & & \Big.-\boldsymbol{{J}}(\boldsymbol{p})_{j=1,k=3}^2\Big),
\end{eqnarray}
\begin{eqnarray}
\mathcal{C}({\tilde{z}})&=&\frac{1}{\text{det}(\boldsymbol{{J}}(\boldsymbol{p}))}\Big(\boldsymbol{{J}}(\boldsymbol{p})_{j=1,k=1}\boldsymbol{{J}}(\boldsymbol{p})_{j=2,k=2} \Big. \nonumber \\ & & \Big. -\boldsymbol{{J}}(\boldsymbol{p})_{j=1,k=2}^2\Big).
\end{eqnarray}
Note that from the above analysis the CRLB is minimized by maximizing the determinant of FIM. Therefore, it is important to manipulate the geometries (position) of anchors to maximize the determinant of FIM. The above analysis is easy to follow for a single node but for multiple nodes in the network, it is possible that changing the position of anchors may reduce the localization error for some nodes but may increase the error for others. Therefore, the optimal position estimation of anchors for multiple nodes in the network can be formulated as a determinant (D)-optimality criteria given as 
\begin{equation}\label{optimalanchors}
\boldsymbol{\breve{B}} = \argmax_{\boldsymbol{\breve{B}}}\sum_{k=1}^{m+n}|\boldsymbol{J}_k|,
\end{equation}
where  $\boldsymbol{\breve{B}} \in \mathbb{R}^{o \times 3}$ are the optimal anchor positions, $k=1,2,...,m+n$ are the number of sensor and relay nodes, and $|\cdot|$ stands for the determinant. Geometrically speaking, the analogue of the above maximization problem for two-dimension networks result in a regular optimal placement of anchors. The regular placement of anchor means that the anchors are placed in such a way that there are no breaking triangles between them for two-dimensional localization. Therefore, here we focus to optimize the depth of anchors to minimize the localization error. To find the optimal depth of anchors, the CRLB for the depth component, $\mathcal{C}({\tilde{z}})$ needs to be minimized which is formulated as
\begin{equation}\label{optimalanchors2}
\boldsymbol{\breve{B}} = \argmin_{\boldsymbol{\breve{B}}}\sum_{k=1}^{m+n}\mathcal{C}({\tilde{z}_k}).
\end{equation}
The optimization problem defined in \eqref{optimalanchors2} is nonconvex and hard to solve. Alternatively, an approximate solution to this problem is achieved by using iterative gradient method as follows
\begin{equation}\label{eq:updaterule}
\boldsymbol{{B}}(t+1)=\boldsymbol{{B}}(t) - \mu^{t}\triangledown_k(\mathcal{C}({\tilde{z}_k(t)}),
\end{equation}
where $t$ is the iteration number, $ \mu^{t}$ is the step size, and $\triangledown_k(\cdot)$  represents the gradient. The selection of step size is crucial for the iterative gradient method and there are several step size selection methods among which the most common is constant step size, i.e., $\mu^{t} = \mu$. However, constant step size leads to poor performance and require a large number of iterations to acquire certain convergence level. Alternatively, the optimal step size ${\mu^{t}}^\star$ can be achieved as
\begin{equation}\label{eq: q}
{\mu^{t}}^\star = \argmin_{\mu} Q\bigg(\boldsymbol{{B}}(t) - \mu\triangledown_k(\mathcal{C}({\tilde{z}_k(t)})\bigg),
\end{equation}
In practice, the minimization problem in \eqref{eq: q} is as hard to solve as the original optimization problem in \eqref{optimalanchors2}. Therefore, a compromise is made between the simplicity of selecting a constant step size and the complexity of  selecting the optimal solution by using Armijo rule \cite{Nocedal2006} given by 
\begin{equation}\label{eq: armigo}
\mu^{t} = \mu \delta^{s},
\end{equation}
where $s \in \{0,1,2,...\}$ and $\delta \in (0,1)$. The anchor positions in \eqref{eq:updaterule} are updated by using the step size calculated in \eqref{eq: armigo}. If $\mathcal{C}(\tilde{z}_k(t+1)) < \mathcal{C}(\tilde{z}_k(t))$ then the anchor positions are updated by \eqref{eq:updaterule} while if  $\mathcal{C}(\tilde{z}_k(t+1)) > \mathcal{C}(\tilde{z}_k(t))$ then the CRLB is not reduced, thus the iterative process stops, and $\boldsymbol{{B}}(t)$ are considered to be the optimal position of anchors for the given network setup.
\section{Numerical Results}\label{results}
In this section, we provide numerical results to validate our proposed solution. In this regard, we consider two scenarios of ten sensor nodes and four relay nodes randomly distributed in a 100 $m^3$ volume as shown in Fig.~\ref{fig:prop_localization}. The transmission range is kept to 80 m to achieve a connected network, ranging error is 0.6 m, and four anchor nodes are considered with their projections at different depths for the global transformation. The performance metric for both of the scenarios is considered to be the root mean square error which is given as
\begin{equation}
\text{RMSE}=\sqrt{\frac{{\parallel\boldsymbol{P}}-\tilde{\boldsymbol{P}}\parallel_F^2}{(m+n)}}. 
\end{equation}
\begin{figure}
\centering
\includegraphics[width=0.7\columnwidth]{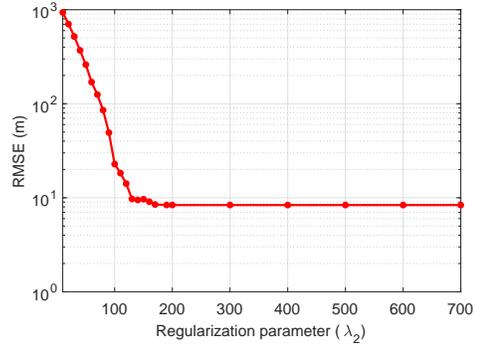}  
\caption{RMSE with respect to $\lambda_2$\label{fig:regularization}} 
\vspace{-1.5 em} 
\end{figure}
In Fig.~\ref{fig:outlierrandom} to Fig~\ref{fig:nooutlieroptimum} the impact of outliers and depth of anchors are investigated. Following four different cases have been evaluated:

\begin{itemize}
\item Case 1: Considering the random depth of anchors in the presence of 35 \% outliers. In this case, the RMSE performance is 23.70 m as shown in  Fig.~\ref{fig:outlierrandom} where the performance is greatly influenced by  both the outliers and random depths of anchors. 

\item Case 2: In this case, we consider that the outliers are still 35 \%  but the depth of anchors is optimized. In this case, the RMSE performance is improved to  8.46 m as shown in  Fig.~\ref{fig:outlieroptimum}.

\item Case 3: In this case, the depth of anchors is random and  the outliers are removed from the estimated pairwise distances. In this case, the RMSE performance is 11.64 m as shown in  Fig.~\ref{fig:nooutlierrandom} where the performance is improved as compared to the first case.

\item Case 4: In this case, the depth of the anchors is optimized as well as the outliers are removed from the estimated pairwise distances. Therefore,  the RMSE performance is greatly improved to 0.659 m as shown in  Fig.~\ref{fig:nooutlieroptimum}. Thus, Fig.~\ref{fig:prop_localization} concludes that removing the outliers from pairwise estimated distances and optimizing the depth of anchors incredibly reduces the localization error for UOWSNs.
\end{itemize}
\begin{figure}
\centering
\includegraphics[width=0.7\columnwidth]{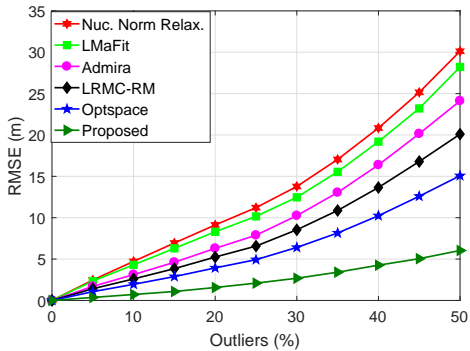}  
\caption{Robustness to outliers\label{fig:outliers}}  
\end{figure}
To analyze the robustness of the proposed 3D localization method in the presence of outliers we consider a scenario of 50 seabed sensor nodes, 4 relay nodes, and 4 anchor nodes with their optimal depths.  A total of 100 Monte Carlo simulation runs were performed and compared with well known matrix completion methods such as nuclear norm minimization, low rank matrix factorization (LMaFit) \cite{ Shen2014}, atomic decomposition for minimum rank approximation (Admira) \cite{ Lee2009}, low rank matrix completion over Riemannian manifold (LRMC-RM) \cite{Bart2013}, and Optspace \cite{ Keshavan2010}. To compare the results, first, we need to find the optimal regularization value $\lambda_2$ which provides the minimum RMSE. Fig.~\ref{fig:regularization} shows the performance of the proposed method in the presence of 35 \% outliers where the optimal regularization values for the given network setup is  $\lambda_2 \geq 150$. Therefore, we set the value of  $\lambda_2 = 150$ for rest of the experiments.
Fig.~\ref{fig:outliers} shows the impact of outliers on the RMSE of the proposed 3D localization method. It is clear from Fig.~\ref{fig:outliers} that the proposed method has better RMSE performance  as compared to the other approaches because it accounts well for the outliers present in matrix $\hat{\boldsymbol{D}}$. 
\begin{figure}
\centering
\includegraphics[width=0.7\columnwidth]{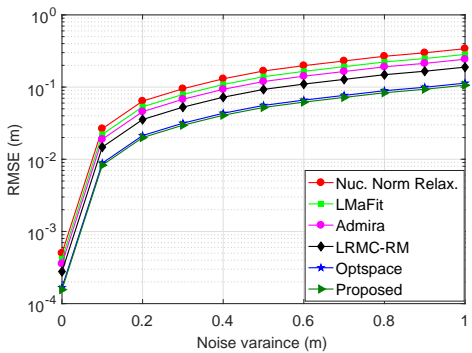}  
\caption{RMSE performace with respect to noise variance\label{fig:noise}} 
\vspace{-1.5em} 
\end{figure}

Undoubtedly, the ranging errors have a negative effect on the accuracy of every localization method. Here, we examine the performance of the proposed technique in the presence of ranging error only. To examine the impact of ranging error we considered 50 seabed sensor nodes, 4 relay nodes, and 4 anchor nodes with their optimal depths. Assuming that the ranging errors are Gaussian distributed with zero mean and variance $\sigma^2$, where the values of $\sigma^2$ are set to 0-1 m. Note that the results are averaged over 100 different network setups. Fig.~ \ref{fig:noise} shows that the proposed localization method have same performance as Optspace in the absence of outliers. 
\begin{figure}
\centering
\includegraphics[width=0.7\columnwidth]{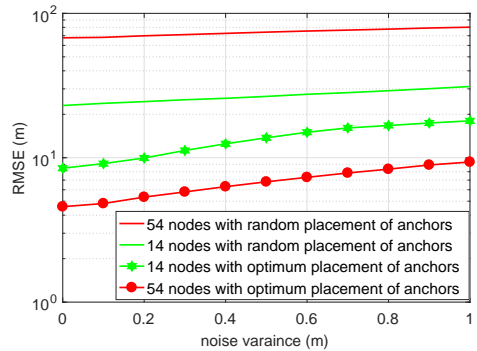}  
\caption{Impact of optimal placement of anchors\label{fig:anchors}} 
\vspace{-1.5em}  
\end{figure}

As mentioned earlier that the proposed 3D localization method optimizes the location of anchors to improve the localization accuracy. Therefore, we have evaluated two different scenarios with 14  and 54 sensor nodes respectively as shown in Fig.~\ref{fig:anchors}. The number of anchors is set to 4 for both setups and RMSE performance is evaluated in terms of noise variance. As illustrated in Fig.~\ref{fig:anchors} that optimizing the depth information of anchors considerably improves the localization performance in both scenarios. Note that the results are averaged our 100 different network setups for both scenarios. Therefore, the results in Fig.~\ref{fig:anchors} conclude that optimizing the depth of anchors in a network significantly improves the localization accuracy.
\section{Conclusions}\label{conc}
Even though two-dimensional localization methods for UOWSNs have been investigated in the past, 3D nature of UOWC environment requires to develop 3D localization methods. Therefore, we have proposed a robust 3D localization method for UOWSNs with limited connectivity.  As the transmission distance of underwater optical sensors is limited, it leads to a partially connected network and many of inter-node distances are missing. Hence, we have employed a low-rank matrix approximation method which can accurately estimate the missing inter-node distances. Additionally, some of the estimated inter-node distances may have a large error and naturally introduces outliers. The traditional 3D network localization methods are susceptible to these outliers. Moreover, the placement of anchors for network localization methods is also an important and challenging problem. Therefore, a closed-form convergent iterative solution is proposed which can accommodate these outliers and optimize the placement of the anchors to improve the localization accuracy. Numerical results validate the performance of the proposed method by showing accurate and robust results to the ranging errors and outliers, respectively.
\section{Acknowledgement}
The authors would like to thank the anonymous reviewers for
their fruitful comments.
\bibliographystyle{../bib/IEEEtran}
\bibliography{../bib/IEEEabrv,../bib/nasir_ref}
\vspace{-1cm}

\begin{IEEEbiography}[{\includegraphics[width=1in,height=1.25in]{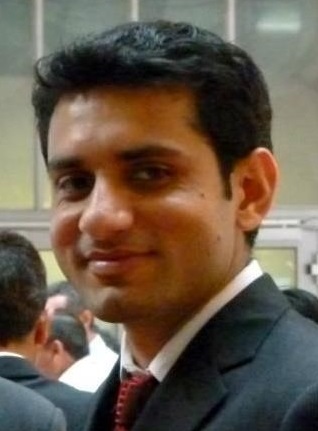}}]{Nasir Saeed}(S'14-M'2016) received his Bachelors of Telecommunication degree from N.W.F.P University of Engineering and Technology, Peshawar, Pakistan, in 2009 and received Masters degree in satellite navigation from Polito di Torino, Italy, in 2012. He received his Ph.D. degree in electronics
and communication engineering from Hanyang University, Seoul, South Korea in 2015. He was an assistant professor at the Department of Electrical Engineering, Gandhara Institute of Science and IT, Peshawar, Pakistan from August 2015 to September 2016. Dr. Saeed worked as an assistant professor at IQRA National University, Peshawar, Pakistan from October 2017 to July 2017. He is currently a postdoctoral research fellow at Communication Theory Lab, King Abdullah University of Science and Technology (KAUST).   His current areas of interest include cognitive radio networks, underwater optical sensor networks, localization, and random matrix theory.
\end{IEEEbiography}

\begin{IEEEbiography}[{\includegraphics[width=1in,height=1.25in]{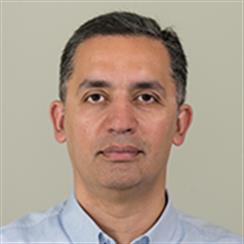}}]{Tareq Y. Al-Naffouri }
(M'10) Tareq Al-Naffouri received the B.S. degrees in mathematics and electrical engineering (with first honors) from King Fahd University of Petroleum and Minerals, Dhahran, Saudi Arabia, the M.S. degree in electrical engineering from the Georgia Institute of Technology, Atlanta, in 1998, and the Ph.D. degree in electrical engineering from Stanford University, Stanford, CA, in 2004. He was a visiting scholar at California Institute of Technology, Pasadena, CA, from January to August 2005 and during summer 2006. He was a Fulbright Scholar at the University of Southern California from February to September 2008. He has held internship positions at NEC Research Labs, Tokyo, Japan, in 1998, Adaptive Systems Lab, University of California at Los Angeles in 1999, National Semiconductor, Santa Clara, CA, in 2001 and 2002, and Beceem Communications Santa Clara, CA, in 2004. He is currently an Associate professor at the Electrical Engineering Department, King Abdullah University of Science and Technology (KAUST). His research interests lie in the areas of sparse, adaptive, and statistical signal processing and their applications and in network information theory. He has over 150 publications in journal and conference proceedings, 9 standard contributions, 10 issued patents, and 6 pending. Dr. Al-Naffouri is the recipient of the IEEE Education Society Chapter Achievement Award in 2008 and Al-Marai Award for innovative research in communication in 2009. Dr. Al-Naffouri has also been serving as an Associate Editor of Transactions on Signal Processing since August 2013.
\end{IEEEbiography}

\begin{IEEEbiography}[{\includegraphics[width=1in,height=1.25in]{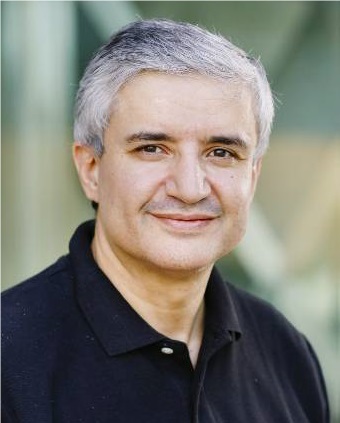}}]{Mohamed-Slim Alouini}
(S'94-M'98-SM'03-F'09) was born in Tunis, Tunisia. He received the Ph.D. degree in Electrical Engineering from the California Institute of Technology (Caltech), Pasadena, CA, USA, in 1998. He served as a faculty member in the University of Minnesota, Minneapolis, MN, USA, then in the Texas A\&M University at Qatar, Education City, Doha, Qatar before joining King Abdullah University of Science and Technology (KAUST), Thuwal, Makkah Province, Saudi Arabia as a Professor of Electrical Engineering in 2009. His current research interests include the modeling, design, and performance analysis of wireless communication systems.
\end{IEEEbiography}

\end{document}